\documentclass[11pt,a4paper]{article}
\usepackage{jheppub}
\usepackage{amsmath}
\usepackage[most]{tcolorbox}
\usepackage{dsfont}
\usepackage{ulem}
\usepackage{natbib}
\usepackage{xcolor}
\usepackage[hang,flushmargin]{footmisc}
\usepackage{tikz-cd}
\usepackage{enumitem}
\usepackage{amsfonts,amssymb, amscd,amsmath,latexsym,amsbsy,bm}
\usepackage{stmaryrd}
\usepackage{todonotes}
\usepackage{float}

\usepackage{romannum}

\makeatletter\renewcommand{\@biblabel}[1]{#1.}\makeatother

\newtcolorbox{empheqboxed}{colback=gray!20, 
 colframe=white,
 width=\textwidth,
 sharpish corners,
 top=0mm, % default value 2mm
 bottom=0pt
}

\title{More solutions to the decoration transformation}
\author{Erdal Catak$^a$ and Mustafa Mullahasanoglu$^{b,c}$}
\affiliation{
$^{a}$ Department of Physics, Istanbul University,\\ 34134 Istanbul, Türkiye \\[-0.4cm]

$^b$ Department of Physics, Bogazici University,\\ 34342 Bebek, Istanbul, Türkiye\\[-0.4cm]

$^c$ Feza Gursey Center for Physics and Mathematics, Bogazici University,\\ 34684, Kandilli,
Istanbul, Türkiye
}

\emailAdd{ecatak@istanbul.edu.tr}
\emailAdd{mustafa.mullahasanoglu@std.bogazici.edu.tr}

\abstract{
In this work, we investigate new solutions to the decoration transformation in terms of various special functions, including the hyperbolic gamma function, the basic hypergeometric function, and the Euler gamma function. These solutions to the symmetry transformation are important to decorate Ising-like integrable lattice spin models obtained via the gauge/YBE correspondence. The integral identities represented as the solution of the decoration transformation are derived from the three-dimensional partition functions and superconformal index for the dual supersymmetric gauge theories. 
}

\keywords{Decoration transformation, dual supersymmetric theories, integrable lattice spin models, dualities.}

\begin{document}

\maketitle

\section{Introduction}

The decoration transformation \cite{Fisher1959, Syozi:1980iw} encodes key symmetry properties of lattice spin models in statistical mechanics. This symmetry transformation allows us to decorate the lattice by adding or removing a spin between interacting spins. The resulting decorated lattice possesses a partition function distinct from the original integrable model up to some coefficient.

In this study, we find the solutions to the decoration transformation for the integrable lattice spin models recently discovered through the gauge/YBE correspondence \cite{Spiridonov:2010em, Yamazaki:2012cp}, see also reviews \cite{Gahramanov:2017ysd, Yamazaki:2018xbx, Yagi:2016oum}. In other words, the Boltzmann weights satisfying the decoration transformation belong to the integrable models in statistical mechanics. These integrable models were investigated in the last decade with the help of the gauge/YBE correspondence.

The gauge/YBE correspondence establishes that partition functions of supersymmetric gauge theories provide solutions to the integrability conditions of lattice spin models. The equality of partition functions or superconformal indices for dual supersymmetric gauge theories yields integral identities. The superconformal indices of certain four-dimensional $\mathcal{N}=1$ supersymmetric dual gauge theories reproduce Spiridonov's elliptic beta integral \cite{Dolan:2008qi}. This framework is well-established, leading to a vast number of integral identities derived from dualities. It was realized that certain such identities can be studied as the star-triangle relation with specific Boltzmann weights, e.g., Spiridonov's integral is the star-triangle relation for the Bazhanov-Sergeev model \cite{Bazhanov:2010kz}. A similar connection has been explored for lens elliptic gamma function solutions \cite{Yamazaki:2013nra, Kels:2017toi, Kels:2017vbc}, which generalize the Bazhanov-Sergeev model to include discrete spins. Thus, the correspondence directly links integral identities to the discovery of new solutions to the star-triangle or star-star relations governing the integrability of lattice spin models. The Boltzmann weights for these integrable models are inherently complicated and are expressed as combinations of special functions.

Building on the gauge/YBE correspondence, it has been shown that other symmetry transformations of the Ising model can also be solved \cite{Mullahasanoglu:2023nes}.  Recent work \cite{Catak:2024ygo} further demonstrates that the decoration transformation can be solved through the equality of partition functions. In this study, the authors focused on three-dimensional $\mathcal{N}=2$ supersymmetric gauge theories, with solutions given by Boltzmann weights expressed in terms of lens hyperbolic hypergeometric gamma functions \cite{Gahramanov:2016ilb, Bozkurt:2020gyy, Catak:2021coz, Mullahasanoglu:2021xyf, Gahramanov:2022jxz}.

Motivated by these lens hyperbolic gamma solutions, we extend the search for solutions to the decoration transformation for the integrable models obtained via gauge/YBE correspondence. We aim to solve the transformation using the Boltzmann weight formulated in terms of the hyperbolic gamma \cite{Spiridonov:2010em}, basic hypergeometric function \cite{Gahramanov:2016wxi, Gahramanov:2013rda, Gahramanov:2014ona, Gahramanov:2015cva, Catak:2022glx}, and complex Euler gamma function \cite{Kels:2013ola, Kels:2015bda, Eren:2019ibl}. We achieve this goal by examining the partition functions of the supersymmetric gauge theories on the squashed sphere $S_b^3$ and the superconformal index on $S^2\times S^1$. The solution involving the complex Euler gamma function is discussed as the dimensional reduction from a three-dimensional squashed lens sphere $S_b^3/\mathbb{Z}_r$.

We highlight that solutions to the decoration transformation are directly applicable to the construction of decorated lattices, whose partition functions differ from the original models by specific multiplicative factors. A good example is the Kagome lattice, which can be derived from the models discussed in this study. The construction of a Kagome lattice from a hexagonal lattice requires only the decoration transformation and the star-triangle relation. We demonstrate that the Boltzmann weights of our integrable models, which already satisfy the star-triangle relation, also provide solutions to the decoration transformation.

The rest of the paper is organized as follows.
In Section 2, we introduce the hyperbolic gamma, the basic hypergeometric function, and the complex Euler gamma solutions of the decoration transformation. For this, we explore corresponding partition functions and superconformal indices of $\mathcal{N}=2$ supersymmetric gauge theories. In the Conclusion, we discuss possible further relations with two sides of the work, which are the lattice spin models and supersymmetric gauge theories. 

%%%%%%%%%%%%%%%%%%%%%%%%%%%%%%%%%%%%%%%%%%%%
\section{The decoration transformation via the gauge/YBE correspondence}
%%%%%%%%%%%%%%%%%%%%%%%%%%%%%%%%%%%%%%%%%%%%
The gauge/YBE correspondence has recently proven to be a powerful connection between lattice spin models and the supersymmetric gauge theories. In the last ten years, the integrability conditions for the star-triangle and the star-star relations have been solved by this bridge. Then, in \cite{Catak:2024ygo}, the correspondence also provides solutions to the other symmetry properties, especially the decoration transformation, of spin models. This transformation maps the integrable model into decorated dual lattice spin models. Partition functions of specific supersymmetric gauge theories yield novel integral identities, which can be interpreted as solutions to the decoration transformation. Consequently, we understood that the dual gauge theories facilitate the construction of dual Ising models. 

In this paper, we will discuss lattice models with $x_i\in \mathbb{R}$ continuous spin variables, in all of them, and  $m_i\in \mathbb{Z}$ discrete spin variables, except the hyperbolic model, represented together as $\sigma_i=(x_i,m_i)\in \mathbb{R}\times \mathbb{Z}$ or in subsets. Only the nearest neighbor spins interact through edges with the Boltzmann weight $W_{\alpha}(\sigma_i,\sigma_j)$ where $\alpha$ is a spectral parameter. All the models we study have self-interactions for each spin variable $S(\sigma_i)$, which are independent of the spectral parameters. 

The partition function of a model can be evaluated in the thermodynamic limit if the Boltzmann weights satisfy the star-triangle relation, which is the simplest version of the Yang-Baxter equation. In this situation, we call the model integrable. Our goal is to determine whether we can decorate these integrable models with the symmetry transformation. We will work on the integrable lattice spin models obtained via the gauge/YBE correspondence, that is, it is shown that all Boltzmann weights of this study satisfy the star-triangle relation.

%We consider decoration -or iteration- transformation \cite{Naya1954, Fisher1959} for lattice spin models in statistical mechanics. The decoration transformation is a map between a spin system consisting of two outer spins interacting with a central spin and a two-spin system with a single interaction. It is a tool to acquire solutions for decorated models since it relates the partition functions of the integrable model and its decorated version up to some coefficient. 

The mathematical expression of the decoration transformation is the following
\begin{align}
   \sum_{m_0} \int dx_0\: S(\sigma_0) W_{\alpha}(\sigma_1,\sigma_0)W_{\beta}(\sigma_2,\sigma_0)
   =\mathcal{R}(\alpha,\beta) W_{\alpha + \beta}(\sigma_1,\sigma_2)
   \label{decorationdefinition},
\end{align}
where, from the left-hand side to the right-hand side, we eliminate the central spin $\sigma_0=(x_0,m_0)$ by integrating the continuous spin variable $x_0$ and summing the discrete spin variable $m_0$. We note that $\mathcal{R}(\alpha,\beta)$ is a spin-independent function which can be normalizable.

%We note that the decoration transformation can be obtained by reducing spins from the star-triangle relation and the reduction is the decrease of the number of flavors from the supersymmetric gauge theories aspect. 

%%%%%%%%%%%%%%%%%%%%%%%%%%%%%%%%%%%%%%%%%%%%%%
\subsection{Hyperbolic solution}
%%%%%%%%%%%%%%%%%%%%%%%%%%%%%%%%%%%%%%%%%%%%%%
Let's first introduce the hyperbolic gamma function \cite{van2007hyperbolic} with the complex variables $\omega_1,\omega_2$
\begin{align}
	\gamma^{(2)}(z;\omega_{1},\omega_{2})=e^{\frac{\pi i}{2}B_{2,2}(z;\omega_{1},\omega_{2})}\frac{(e^{-2\pi i\frac{z}{\omega_{2}}}\tilde{q};\tilde{q})_\infty}{(e^{-2\pi i\frac{z}{\omega_{1}}};q)_\infty} \; ,
\end{align}
where $\tilde{q}=e^{2\pi i \omega_{1}/\omega_{2}}$, $q=e^{-2\pi i \omega_{2}/\omega_{1}}$
and the Bernoulli polynomial is 
$
 B_{2,2}(z;\omega_1,\omega_2)=\frac{z^2-z\omega}{\omega_1\omega_2}+\frac{\omega^2+\omega_1\omega_2}{6\omega_1\omega_2}$, and  $\omega:=\omega_1+\omega_2$.

We use the $q$-Pochhammer symbol $(z;q)_{\infty}=\prod_{i=0}^{\infty}(1-zq^i)$ with the notation $(z,x;q)_\infty=(z;q)_\infty(x;q)_\infty$. The hyperbolic gamma function has the reflection property 
\begin{align}
  \gamma^{(2)}(\omega_1+\omega_2- z ;\omega_1,\omega_2)\gamma^{(2)}( z ;\omega_1,\omega_2)=1 \:.\label{hypref}
\end{align}
We solve the decoration transformation using a hyperbolic beta integral of Askey-Wilson type \cite{STOKMAN2005119, Ruijsenaars2003, Kels:2018xge, Sarkissian:2020ipg}.

The partition functions of the $\mathcal{N} = 2$ supersymmetric dual gauge theories on the squashed three-sphere give the hyperbolic integral identities \cite{Hosomichi:2014hja, Willett:2016adv}. In this study, we have two dual theories, which are called Electric and Magnetic historically.  In the Electric theory, there are the $SU(2)$ gauge symmetries and $N_f=4$ flavors where chiral multiplets transform under the fundamental representation of the gauge group and the flavor group, and the vector multiplet transforms as the adjoint representation of the gauge group. The Magnetic theory does not involve gauge symmetry, and it consists of six mesonic operators together with a singlet chiral multiplet in the totally antisymmetric tensor representation of the flavor group. This duality is a reduced version of the dual theories with the $SU(6)$ flavor symmetry. So, we have the $SU(4)$ flavor symmetry, and fifteen chiral multiplets are reduced to the seven chiral multiplets in the magnetic theory. 

The partition functions of the dual theories with the $SU(6)$ flavor symmetry give a solution to the star-triangle relation in \cite{Spiridonov:2010em}. For this integrable model, we look for the decoration transformation and solve it with the same Boltzmann weights. To reach this, we focus on the partition functions of dual gauge theories with $SU(4)$ flavor symmetry and its following result 
\begin{align}
    \int_{-i\infty}^{i\infty}&\frac{\prod_{j=1}^4\gamma^{(2)}(a_j\pm z;\omega_1,\omega_2)}{\gamma^{(2)}(\pm2z;\omega_1,\omega_2)}\frac{dz}{2i\sqrt{\omega_1\omega_2}}
     = \frac{\prod_{1\leq i<j\leq 4}\gamma^{(2)}(a_i+a_j;\omega_1,\omega_2)}{\gamma^{(2)}(\sum_{i=1}^4a_i;\omega_1,\omega_2)}
     \:,
    \label{intssr}
\end{align}
where the shorthand notation is
\begin{align}
  \gamma^{(2)}(\pm z ;\omega_1,\omega_2)=\gamma^{(2)}(+ z ;\omega_1,\omega_2)\gamma^{(2)}(- z ;\omega_1,\omega_2) \:.
\end{align}
%Note that there is no balancing condition in (\ref{intssr}) and in the integral identities studied in the rest of the paper. Also, the mathematical structures of the integral identities, which we will not go into detail, can be found in \cite{BultThesis} and references therein.
The integral identity (\ref{intssr}) is discussed as solutions to the star-triangle relation in \cite{Kels:2018xge} with different types of Boltzmann weights and solution to the star-square relation \cite{Mullahasanoglu:2023nes}. However, we will show that the same integral identity is a decoration transformation for the Boltzmann weights of the integrable hyperbolic model \cite{Spiridonov:2010em}  
 \begin{align}
    \begin{aligned}
W_\alpha(x_i,x_j)=\gamma^{(2)}(-\alpha \pm x_i\pm x_j ;\omega_1,\omega_2)\:,
\end{aligned}\label{integrableB1}
\end{align}
under the change of the variables applied to (\ref{intssr})
\begin{align}
    \begin{array}{c}
 a_j=\pm x_1-\alpha\,,\quad 
  j=1,2\:, \\
   a_j=\pm x_2-\beta \,,\quad 
  j=3,4\:.
\end{array}
\label{chngvs}
\end{align}
$z$ is relabeled as a central continuous spin $x_0$, and we obtain the vector multiplet as a self-interaction term 
\begin{align}
   S(x_0)= \frac{1}{\gamma^{(2)}(\pm2x_0;\omega_1,\omega_2)}\:.\label{self1}
\end{align}
We also have a spin-independent function 
\begin{align}
    \mathcal{R}(\alpha,\beta)=\frac{\gamma^{(2)}(-2\alpha ;\omega_1,\omega_2)\gamma^{(2)}(-2\beta ;\omega_1,\omega_2)}{\gamma^{(2)}(-2(\alpha+\beta) ;\omega_1,\omega_2)}\:.
\end{align}
With these definitions (\ref{intssr}) turns into the decoration transformation (\ref{decorationdefinition}) without discrete spin variables. This means that we can decorate and construct various types of lattice models from integrable models by adding extra spins to the system with the decoration transformation. 
%%%%%%%%%%%%%%%%%%%%%%%%%%%%%%%%%%%%%%%%%
\subsection{Trigonometric solution}
%%%%%%%%%%%%%%%%%%%%%%%%%%%%%%%%%%%%%%%%%
The trigonometric lattice spin model is obtained by solving the star-triangle relation using the basic hypergeometric integral identities \cite{ Gahramanov:2015cva, Gahramanov:2023lwk}. On the gauge side, these identities stand for the partition functions of the dual gauge theories on $S^2 \times S^1$. The dual Electric and Magnetic theories of this case have the same ingredients, just like the duality mentioned above, studied on the squashed three-sphere. 

The integral identity in which the dual theories consist of the $SU(6)$ flavor symmetry is studied as a solution to the star-triangle relation.  The partition functions of the dual theories with $SU(4)$ flavor symmetries are explicitly evaluated in \cite{Gahramanov:2016wxi}, and the integral identity is obtained
\begin{align} \nonumber
& \sum_{y\in\mathbb{Z}}\oint \frac{dz}{4 \pi i z}\frac{(1-q^{y} z^2)(1-q^{y} z^{-2})}{q^y     z^{4y} }
\prod_{j=1}^4 
\frac{(q^{1+\frac{u_j+y}{2} }/{a_jz},q^{1+\frac{u_j-y}{2} }{z}/{a_j};q)_\infty}
{(q^{\frac{u_j+y}{2} }a_jz,q^{\frac{u_j-y}{2} }{a_j}/{z};q)_\infty}
  \\ 
&\quad= \frac{( q^{\frac{\sum_{i=1}^4 u_i}{2}}a_1 a_2 a_3 a_4;q)_{\infty}}{(q^{1+ \frac{\sum_{i=1}^4 u_i}{2}}/a_1 a_2 a_3 a_4;q)_{\infty}} \prod_{1\leq j<k\leq 4}  \frac{(q^{1+\frac{u_j+u_k}{2}}/a_ja_k;q)_\infty}
{(q^{\frac{u_j+u_k}{2}}a_ja_k;q)_\infty}\:.
\label{s2s1int}
\end{align}

To obtain the Boltzmann weight of the integrable trigonometric model 
\begin{align}
 \begin{aligned}
    W_{\alpha, \beta}(\sigma_i,\sigma_0)=
    \frac{(q^{1+(-\beta_i\pm v_i\pm v_0)/2}(\alpha_i^{-1}x_i^{\pm1}x_0^{\pm1})^{-1};q)_\infty}{(q^{(-\beta_i\pm v_i\pm v_0)/2}\alpha_i^{-1} x_i^{\pm1}x_0^{\pm1};q)_\infty}\;.
    \end{aligned}\label{integrableB3}
\end{align}
We redefine fugacities as
\begin{align}
 a_j=x_1^{\pm1}\alpha_1^{-1}\quad 
  j=1,2\:, \quad \quad  u_j=\pm m_1-\beta_1\quad 
  j=1,2\:, \\
   a_j=x_2^{\pm1}\alpha_2{-1} \quad 
  j=3,4\:, \quad \quad u_j=\pm m_2-\beta_2\quad 
  j=3,4\:.
\end{align}
The self-interaction term and the spin-independent function appear 
\begin{align}
    S(\sigma_0)=\frac{(1-q^{m_0} x_0^2)(1-q^{m_0} x_0^{-2})}{q^{m_0}     x_0^{4m_0} }\:, \quad \quad    \mathcal{R}(\alpha,\beta)=\frac{\prod_{i=1}^2\frac{(q^{1-\beta_i}\alpha_i^{2};q)_\infty}{(q^{-\beta_i}\alpha_i^{-2} ;q)_\infty}}{\frac{(q^{1-\beta_1-\beta_2}\alpha_1^{2}\alpha_2^{2};q)_\infty}{(q^{-\beta_1-\beta_2}\alpha_1^{-2}\alpha_2^{-2} ;q)_\infty}}
    \;.\label{self3}
\end{align}
With these definitions, the basic hypergeometric integral identity (\ref{s2s1int}) takes the form of the decoration transformation (\ref{decorationdefinition}).

%%%%%%%%%%%%%%%%%%%%%%%%%%%%%%
\subsection{Rational solution}
%%%%%%%%%%%%%%%%%%%%%%%%%%%%%%
In this part, we obtain the decoration transformation from the lens hyperbolic solution. The dual theories with the same matter content live on the three-dimensional squashed lens space $S^3_b/\mathbb{Z}_r$. Ordinary limit of the hyperbolic hypergeometric solution of the decoration transformation $r\to \infty$ \cite{Benini:2011nc, Benini:2012ui}, see {\it The Integral Identity II} in \cite{Mullahasanoglu:2024stv} for more details, gives us a general complex analogue of the de Branges-Wilson integral \cite{Sarkissian:2020ipg, andrews_askey_roy_1999}
\begin{align}\begin{aligned}
\sum_{y\in {\mathbb Z}}&\int_{-\infty}^{\infty}\big(z^2+y^2\big)\prod_{i=1}^4{\bf \Gamma}(a_i\pm z,u_i\pm y)\frac{dz}{ 8\pi}
=\frac{\prod\limits_{1\leq i < j \leq 4}
{\bf \Gamma}(a_i+a_j,u_i+u_j)}{{\bf \Gamma}\Big(\sum\limits_{i=1}^ 4 a_i,\sum\limits_{i=1}^ 4 u_i\Big)}
\:,
\label{hyper44}
\end{aligned}\end{align} 
where the complex gamma function\footnote{Lens hyperbolic gamma function \cite{Gahramanov:2016ilb} is defined with $\text{Im}(\omega_1/\omega_2)>0$ 
\begin{align}
\gamma_h(z,y;\omega_1,\omega_2) 
=    \gamma^{(2)}(-iz-i\omega_1y;-i\omega_1r,-i\omega) \times \gamma^{(2)}(-iz-i\omega_2(r-y);-i\omega_2r,-i\omega) \:,
\end{align}
where $r\in\{1,2,...\}$, $y\in \{0,1,...,r-1\}$. The ordinary limit is the following
\begin{equation}
	\lim_{r\to\infty} \gamma_h(z,y;\omega_1,\omega_2)=\Big(\frac{r}{4\pi}\Big)^{\frac{2-2iz}{r}}\frac{\Gamma\left(\frac{iz+y}{2}\right)}{\Gamma\left(1-\frac{iz-y}{2}\right)}\equiv \Big(\frac{r}{4\pi}\Big)^{\frac{2-2iz}{r}}{\bf \Gamma}(z,y)\:.
\end{equation}} is a certain combination of Euler's gamma functions
\begin{equation}
{\bf \Gamma}(x,n)
=\frac{\Gamma\big(\frac{n+{i}x}{2}\big)}{\Gamma\big(1+\frac{n-{ i}x}{2}\big)},
\label{Cgamma}\end{equation}
where $x\in {\mathbb C}$ and $n\in {\mathbb Z}$.
The reflection property is
\begin{equation}
{\bf \Gamma}(x,-n)=(-1)^n{\bf \Gamma}(x,n)\:,
\quad \&\quad
{\bf \Gamma}(x,n){\bf \Gamma}(-x-2{ i},n)=1\:.
\label{gamma}\end{equation}
The star-triangle relation for the following Boltzmann weight is solved in \cite{Kels:2013ola, Kels:2015bda, Eren:2019ibl} 
 \begin{align}
    \begin{aligned}
        W_{\alpha, \beta}(\sigma_i,\sigma_0)={\bf \Gamma}(-\alpha \pm x_i\pm x_0,-\beta \pm v_i\pm v_0 )
       \:. \label{integrableB4}
    \end{aligned}
\end{align}
We change the variables in (\ref{hyper44}) to show that we can solve the decoration transformation for this  Boltzmann weight (\ref{integrableB4})
\begin{align}
 a_j=\pm x_1-\alpha_1\quad 
  j=1,2\:, \quad \quad  u_j=\pm m_1-\beta_1\quad 
  j=1,2\:, \\
   a_j=\pm x_2-\alpha_2 \quad 
  j=3,4\:, \quad \quad u_j=\pm m_2-\beta_2\quad 
  j=3,4\:.
\end{align}
The self-interaction term and the spin-independent function are
\begin{align}
    S(\sigma_0)=x_0^2+v_0^2 \quad \quad \mathcal{R}(\alpha,\beta)=\frac{\prod_{i=1}^2{\bf \Gamma}(-2\alpha_i ,-2\beta_i )}{{\bf \Gamma}(-2(\alpha_1+\alpha_2) ,-2(\beta_1+\beta_2) )}\:.\label{self4}
\end{align}

\section{Conclusion}

It is discovered that the symmetry transformations of the Ising-like models can be solved with the integral identities coming from partition functions of the dual supersymmetric gauge theories. In \cite{Catak:2024ygo}, a lens hyperbolic solution to decoration transformation is presented for the corresponding integrable model obtained via the gauge/YBE correspondence. 

In this paper, we achieved solving the decoration transformation with the Boltzmann weights of the hyperbolic model, the trigonometric model, and the rational model. In general, the equality of the partition functions for the theory with $SU(2)$ gauge symmetry and $SU(4)$ flavor symmetry and its dual theory with the same global symmetry without gauge degrees of freedom gives the integral identity. 

There are various directions and mathematical connections to the integral identities solved as integrability conditions, see \cite{Gahramanov:2022qge} and references therein. Therefore, it would be interesting to see the role of the decoration transformation discussed in this study.

\section*{Acknowledgements}
We are grateful to Ilmar Gahramanov for his valuable comments on the manuscript.
We are supported by the Istanbul Integrability and Stringy Topics Initiative (\href{https://istringy.org/}{istringy.org}). Erdal Catak and Mustafa Mullahasanoglu are supported by the 3501-TUBITAK Career Development Program under grant number 122F451.

%\section*{Data Availability Statement}
%No Data associated in the manuscript.

%\appendix
%\section{Ising model and relations}

\bibliographystyle{utphys}
\bibliography{refYBE}

\end{document}